\documentclass[aps,showpacs,prl,toolkits,twocolumn]{revtex4}
\newcommand{\dslash}{D\!\!\!\!\slash}
\begin{document}

\title {QCD inequalities for the nucleon mass and the free energy of baryonic matter }

\author{Thomas D. Cohen}
\email{cohen@physics.umd.edu}

\affiliation{Department of Physics, University of Maryland,
College Park, MD 20742-4111}

\begin{abstract}
The positivity of the integrand  of certain Euclidean space
functional integrals for two flavor QCD with degenerate quark masses  implies  that the free energy per unit
volume for QCD with a baryon chemical potential $\mu_B$ (and zero
isospin chemical potential) is greater than the free energy with
isospin chemical potential $\mu_I = \frac{2 \mu_B}{N_c} $ (and zero
baryon chemical potential).  The same result applies to QCD with any number of heavy flavors in addition to the two light flavors so long as the chemical potential is understood as applying to the light quark contributions to the baryon number. This relation implies a bound on the nucleon mass: there exists a particle $X$ in QCD (presumably the pion) such that  $M_N \ge \frac{N_c \, m_X }{2 \, I_X}$ where $m_X$ is the  mass and of the particle and $I_X$ is its isospin.
\end{abstract}


\maketitle

Quantum chromodynamics (QCD) is the theory underlying strong interactions.  The theory
is not analytically tractable via perturbative and other weak
coupling methods except for a limited set of observables in a limited kinematic regime.  While
certain aspects of the nonperturbative regime of QCD have been
explored via lattice QCD\cite{lat}, it is important to 
establish as many properties of QCD as
possible via rigorous analytic means.  QCD inequalities  represent an ideal method to do this.  An early variant of the approach was based on a demonstration by Nussinov\cite{Nus83} that bounds could be placed on hadronic quantities in a wide class of models inspired by QCD.  The approach was greatly strengthened by the realization of
Weingarten \cite{Wei83} and Witten \cite{Wit83} that similar bounds could be obtained directly from QCD itself through inequalities based on Eulcidean space functional integral representations of physical quantities. Thus, the approach
is a method to deduce certain
qualitative features of QCD from first principles. While the resulting inequalities have not been proved as theorems at the level of rigor demanded by mathematicians, they make use of only the most standard assumptions made by physicists.  For example one assumes that the theory exists, that physical quantities may be represented via  functional integrals, that a Wick rotation from Minkowski space to Euclidean space is innocuous and so on.  The field of QCD inequalities is now two decades old; the state of the art is presented in a recent comprehensive review by Nussinov and Lampert~\cite{LamNus00}. Although the results of QCD inequalities
are qualitative, they can serve to supplement understanding gleaned
from lattice QCD studies; they provide an analytic means of
understanding some of the results of QCD which are both seen in
nature and which emerge from numerical studies on the lattice.
QCD inequalities may also provide insight into certain properties of QCD which are not tractable on the lattice using  Monte Carlo algorithms.

The underlying idea of QCD inequalities is quite simple.  One
relates a physical quantity of interest to a Euclidean space
functional integral over gauge field configurations.  If
a second quantity can be represented via a functional integral for
which the integrand is greater than or equal to the integrand of
the initial quantity  for all gauge configurations, then one
can conclude that the second functional integral is bigger than
the first and this in turn allows one to bound one physical
quantity by another. 

This letter focuses on the use of QCD inequalities in
two apparently unrelated problems in strong
interaction physics.  The first is bounding the mass of the
nucleon from below in  terms of other physical observables.  This is an old problem. Nussinov derived a bound that the nucleon mass must be greater than or equal to $\frac{3 m_\pi}{2}$ in the context of QCD-inspired models.  Weingarten, in his original paper, 
attempted to bound the mass of the nucleon  as some multiple
of the pion mass directly from QCD \cite{Wei83}.  Unfortunately, the only rigorous method found required the study of QCD with six or more degenerate light flavors.  An alternative approach was also suggested by Weingarten; however, it used plausible but not provable assumptions about bounds on the quark propagator in the presence of an arbitrary gauge potential.  Subsequently,  Nussinov and Sathiapalan \cite{NusSat85}  showed that in the large $N_c$ limit of QCD with two degenerate flavors, the nucleon mass is bounded by $ M_N > \frac{N_c \, m_\pi}{2}
$.  However, previously no rigorous lower bound has been obtained for the nucleon mass directly from QCD at finite $N_c$ for two degenerate flavors.  Such a bound is derived here.

The second issue discussed in this letter is the problem of finding a lower bound on the free energy of QCD at nonzero
baryon chemical potential, or more precisely a baryon chemical potential associated with the two light degenerate flavors. This second problem is significant
because for the zero temperature case it is in the class of 
problems for which Monte Carlo algorithms cannot be used in lattice simulations.  Moreover no viable alternative presently exists for doing such simulations.  Thus any reliable results from the theory are extremely welcome.  The problem is also interesting in light of the intense recent interest in QCD at
finite baryon density \cite{rho}.

The bound on the QCD free energy at
fixed baryon chemical potential will play an essential role in
bounding the nucleon mass and accordingly this problem will be
treated first.  This problem is rather unusual for a QCD
inequality treatment.  The approach is more commonly associated with
bounding masses of particles via the study of correlation functions rather than with thermodynamically intensive quantities such as free energy densities. However, QCD inequalities have been used for intensive quantities in the past.  Vafa and Witten \cite{VafWit84} demonstrated that the
vacuum energy for QCD including a $\theta$ term has an absolute
minimum at $\theta=0$.  While the validity of a related argument by Vafa and Witten\cite{VafWit84} that parity cannot be spontaneously broken has recently been questioned \cite{questions}, the demonstration that the the minimum vacuum energy is at $\theta=0$ is clearly correct.  

The  Vafa-Witten proof is very simple.
The integrand of the functional integral is given by
$\prod_{i={\rm flavors}} \, {\rm det}(\dslash + m_i) \,
e^{-S_{YM} \, + \, i \theta \nu }$ where $\, {\rm det}(\dslash + m_i)$, the functional determinant for a given flavor is real and
non-negative \cite{Wei83},  $S_{YM}$ is the Yang-Mills action, and
$\nu$ is the winding number.  Thus, the only effect of setting
$\theta$ to be nonzero is to multiply the rest of the integrand
(which is real and positive) by a pure phase factor $e^{i \theta
\nu }$.  Since the real part of this phase factor is always less
than or equal to unity (and the imaginary part will integrate to
zero) one immediate deduces that the functional integral for
nonzero $\theta$ is bounded from above by the integral with
$\theta=0$. Finally, identifying the functional integral as the
generating function $Z(\theta) = e^{- V{\cal E}(\theta)}$ where
$V$ is the four dimensional volume and ${\cal E}(\theta)$ is
 the vacuum energy as a function of $\theta$, one sees that the
 inequality for the generating
 function implies that ${\cal E}(\theta) > {\cal E}(0)$.

Here an argument analogous to that of Vafa and Witten will be
given for the problem of two flavor QCD with degenerate quark masses  at a nonzero chemical potential.
The free energy density for QCD at fixed temperature and baryon
chemical potential, ${\cal G}_B(T,\mu_B)$ is given in terms of the grand partition function $Z_B(T, \mu_B)$ as ${\cal G}_B(T,\mu_B)
\, = - \,(\beta V_3)^{-1} \log \left( Z_B(T, \mu_B) \right )$ where
$V_3$ is the three dimensional volume of the system and $\beta$
is the inverse temperature. In QCD with two degenerate flavors,
$Z_B(T,\mu_B)$ can be represented as a functional integral,
\begin{eqnarray}
&~&Z_B(T,\mu_B) \, =  \nonumber \\
&~&\int  d [A] \, \left ( \, {\rm det}(\dslash + m - \frac{\mu_B}{N_c} \,
\gamma_0 \, ) \, \right )^2 e^{-S_{YM}}
\end{eqnarray}
where $N_c$ is the number of colors (3 for the physical world),
the functional determinant is over a single quark flavor and the
temperature is implemented by imposing periodic boundary
conditions in time for the gluon fields $A(t+\beta)=A(t)$ with
$\beta=1/T$; similarly antiperiodic boundary conditions for the
fermions are imposed in the functional determinant.  The factor
of $\frac{1}{N_c}$ simply reflects the fact that the chemical potential
is for baryon number and the baryon number of a single quark is
$\frac{1}{N_c}$.  The difficulty in simulating this functional integral
on the lattice stems from the fact that the functional
determinant is not generally real and positive.  From the
perspective of QCD inequalities, however, this is not a bug, but a
feature; it allows one to bound the partition function from
above:
\begin{eqnarray}
&~& Z_B(T,\mu_B) \le \nonumber \\
&~&\int  d [A] \left | \, {\rm det}(\dslash + m - \frac{\mu_B}{N_c}  \,
\gamma_0 ) \right |^2 e^{-S_{YM}} \; \; . \label{partitionineq}
\end{eqnarray}

In order for inequality (\ref{partitionineq}) to be useful, its right-hand side needs to be expressed in terms of a physical quantity.
Fortunately it can be related to the free energy density of QCD
with an {\it isospin} chemical potential \cite{AlfKapWil99}.  An isospin
chemical potential term is of the form $\mu_I \, \overline{q}
\gamma_0 \frac{\tau_3}{2} q$,  which implies that the functional integral
for the grand partition function $Z_I(T, \mu_I) = \exp \left(
{- \beta \, V_3 \, \cal G}_I(T,\mu_I) \right)$ is given by
\begin{eqnarray}
&~& Z_I(T,\mu_I) =  \int d [A] \, e^{-S_{YM}} \nonumber \\
&\times& \, {\rm det}(\dslash + m -
\frac{\mu_I}{2} \, \gamma_0) \, {\rm det}(\dslash + m + \frac{\mu_I}{2}\,
\gamma_0) ~~~~~ \label{ZI}
\end{eqnarray}
where the two
functional determinants are for the two flavors of quark and the
opposite signs of the $\mu_I$ terms reflect the opposite values of
$I_3$ for the two flavors.  To proceed we use the fact that
\begin{eqnarray}
\gamma_5 (\dslash + m + \frac{\mu_I}{2}  \gamma_0) \gamma_5 & = &(-\dslash
+ m - \frac{\mu_I}{2} \,
 \gamma_0)  ~~~~\nonumber \\
& = &
(\dslash + m - \frac{\mu_I}{2}\,  \gamma_0)^{\dagger} \label{gammaprop}
\end{eqnarray}
where the last equality exploits the fact that in Euclidean space
$\dslash$ is anti-Hermitian while the other two operators are
Hermitian.  Exploiting the cyclic property of the determinant
allows one to write the second functional determinant
in eq.~(\ref{ZI}) as
$\, {\rm det}(\dslash + m + \frac{\mu_I}{2}\,  \gamma_0) = \, {\rm
det}\left(\gamma_5 (\dslash + m + \frac{\mu_I}{2}\,  \gamma_0)
 \gamma_5 \right )$ and using equation
eq.~(\ref{gammaprop}) then gives
\begin{eqnarray}
&~&{\rm det}(\dslash + m + \frac{\mu_I}{2}\,  \gamma_0) = \nonumber \\
&~&\left ( \, {\rm det}(\dslash + m - \frac{\mu_I}{2}\,  \gamma_0)
\right )^*\label{CC}
\end{eqnarray}
Combining eq.~( \ref{CC}) with eq.~(\ref{ZI}) yields
\begin{eqnarray}
&~& Z_I(T,\mu_I) = \nonumber \\
&~&\int  d [A] \left | \, {\rm det} (  \dslash + m - \frac{\mu_I}{2}  \,
\gamma_0 \, ) \, \right |^2 e^{-S_{YM}} \; \; . \label{ZI2}
\end{eqnarray}
Finally, inequality (\ref{partitionineq}) together with
eq.~(\ref{ZI2}) implies  that $Z_I(T,\frac{2 \mu_B}{N_c}
) \ge
Z_B(T,\mu_B)$. This relation together with the definition of the
free energy requires that
\begin{equation}
{\cal G}_B(T,\mu_B) \ge {\cal G}_I(T, \frac{2 \mu_B}{ N_c}) \,
\label{ineq1}
\end{equation}
which is the first principal result of this letter.

Although this result was derived for two flavor QCD, the argument goes through for QCD with two degenerate light flavors and additional heavy flavors, provided the chemical potential term is understood as being the chemical potential for the up and down quark contributions to the baryon number rather than the full baryon chemical potential. 
 The only change in the argument needed for this more general case is to include the functional determinant for the heavy flavors in all of the functional integrals. As the chemical potential does not apply to these heavy flavors,  the functional determinants are the same as at $\mu=0$ and hence are real and non-negative. Thus, they do not alter the preceding inequalities. This more general case is significant as in nature QCD has two light quarks which are nearly degenerate and additional heavy flavors.
The result also applies in the general case to the full baryon chemical potential if one is in a regime in which the $\overline{s} \gamma_0 s = \overline{c} \gamma_0 c =\overline{b} \gamma_0 b = \overline{t} \gamma_0 t =0$, since in this regime the total baryon  number comes from up and down quarks.   Such a regime occurs at zero temperature provided the chemical potential is below the critical chemical potential for strangeness condensation to occur.

A bound on the nucleon mass may be derived from  inequality (\ref{ineq1}) using thermodynamic
arguments. The bound applies to QCD with two degenerate light flavors and any number of heavy flavors. To begin, note that inequality (\ref{ineq1}) holds at
all temperatures, including $T=0$. At zero temperature there are
no thermal fluctuations and the system is in a single quantum state
 which  minimizes the free energy $G=H-\mu N$, where $\mu$ is the relevant chemical potential (either isospin or baryon)
and $N=V_3 \rho$ is the related particle number.  The role of the chemical potential at zero temperature is simply to alter the relative free energies of the various quantum states.  Thus
increasing the chemical potential from zero at $T=0$ will have no
effect until it is large enough so that the free energy of some
other quantum state drops enough to equal that of the true
vacuum.  Accordingly, there is a critical value for the absolute
value of the chemical potential at $T=0$ below which the density is zero
and the free energy is that of the vacuum state (which is
conventionally taken to be zero). The critical chemical potentials
are thus defined as follows:
\begin{eqnarray}
{\cal G}_B(T=0,\mu_B) &  = &  0\;
\;{\rm for} \; \;|\mu_B| < \mu_B^c \; ,\nonumber \\
{\cal G}_B(T=0,\mu_B) &  < & 0\;
 \; {\rm for} \; \; |\mu_B| > \mu_B^c \; , \nonumber \\
{\cal G}_I(T=0,\mu_I) &  = & 0\;
\; {\rm for} \; \; |\mu_I| < \mu_I^c \; ,\nonumber \\
{\cal G}_I(T=0,\mu_I) &  < & 0 \;
 \; {\rm for} \; \; |\mu_I| > \mu_I^c \;.
\label{crit}
\end{eqnarray}
Inequalities (\ref{ineq1}) and (\ref{crit}), 
 together imply the  relation
\begin{equation}
\mu_B^c \ge \frac{N_c \, \mu_I^c}{2} \label{ineq2}
\end{equation}

It is straightforward to bound the $\mu_B^c$ from above by
the nucleon mass using a simple variational argument.  Consider
the quantum state of a single nucleon at rest.  This state has
energy $M_N$ and baryon number unity; its free energy is $G_B =
M_N - \mu_B$.  Clearly this is less than or equal to zero for
$\mu_B \ge M_N$.  Thus, there is at least one state lower than
the vacuum whenever $\mu_B \ge M_N$.  This implies that $\mu_B^c
\le M_N$.  This last inequality would become an equality if the
system for $\mu_B$ just above its critical value formed an
arbitrarily low density gas of nucleons (implying a second order
transition).  However, this is not what actually happens.
Based on the
extrapolation of finite nuclei densities and masses to infinite
nuclear matter \cite{nucmat} one has a solid empirical basis to
conclude that the transition is first order:
for $\mu_B$ just above $\mu_B^c$ the system has
nonzero energy and nonzero density.  Thus $\mu_B^c = M_N - B$
where $B$ is the binding energy per nucleon of infinite nuclear
matter and the inequality, $\mu_B \ge M_N$, is not saturated. In
any event, the inequality $\mu_B^c
\le M_N$ together with inequality (\ref{ineq2})
yields a bound on the nucleon mass.
\begin{equation}
M_N \ge \frac{N_c \,  \mu_I^c}{2} \; . \label{ineq3}
\end{equation}

For inequality (\ref{ineq3}) to be useful we need to know 
$\mu_I^c$.   For sufficiently
small values of the quark mass we know that chiral perturbation
theory accurately describes low energy excitations of the QCD
vacuum.  In chiral perturbation theory, the phase transition
associated with increasing $\mu_I$ is second order and amounts to pion condensation, which implies $\mu_I^c = m_\pi$  \cite{SonSte}.
More generally, $\mu_I^c$ corresponds to the state in QCD with the lowest energy per unit isospin.  Let us denote such a state as $X$.  There are two possibilities for $X$.  If the transition is second order (as it is in chiral perturbation theory), then $X$ is a single particle state; such a state is clearly at zero momentum  so the energy can be identified as the mass of the particle.  In this case, we can denote the mass of the state $X$, as $m_X$, and its isospin as , $I_X$. The inequality (\ref{ineq3}) becomes
\begin{equation}
M_N \ge \frac{N_c \, m_X}{ 2 \, I_X }  \; . \label{ineq4}
\end{equation}
 
 Although it is strongly believed that in nature the transition is  second order, we do not require this in obtaining an inequality. Assume for the moment that the transition were first order.  In this case the state of minimum energy per isospin would in fact be infinite isospin matter (in analogy to infinite nuclear matter), {\it i.e.} a state of uniform isospin density and uniform energy density filling all space.  However, if this were the case, one could construct single particle states by taking large chunks of isospin matter (in analogy to large nuclei in a world where the electro-magnetic interaction was shut off).  The energy of such a state will have a bulk contribution scaling with the volume of the chunk and equaling $\mu_I^c I_3$ where $I_3$ is the isospin of the state.  Corrections to this bulk value scale as the surface area and hence go like $I_3^{2/3}$.   Thus, by making $I_3$ arbitrarily large but finite one can can create  single particle states whose energy/isospin is arbitrary close $\mu_I^c$. Identifying such a state as $X$ one finds that there exists an $X$ for which $\frac{m_X}{I_X}$ can be made arbitrarily close to $\mu_I^c$.  Accordingly, regardless of whether the transition is first or second order one can conclude that inequality (\ref{ineq3}) implies there exists in QCD some single particle state $X$ such that inequality (\ref{ineq4}) holds.   

Note, the state $X$ cannot be the nucleon itself as this is inconsistent with the inequality above; hence, inequality (\ref{ineq4}) makes a nontrivial prediction.

A few comments on the significance of these
results is in order.  Consider the
implications of inequality (\ref{ineq1}).  As noted previously,
QCD at finite baryon chemical potential cannot be simulated on the lattice
using Monte Carlo methods. It is generally believed that
asymptotically high densities can be treated analytically  
from QCD using weak coupling, but nonperturbative, 
techniques along the line of BCS theory \cite{Son,rho}.  However, it
is also generally believed that these densities are
extraordinarily high and the regime where these methods work is
unlikely to be relevant  in laboratory experiments or
in stellar physics.  Thus,
studies of phenomenological significance have been based on various models \cite{model,rho}.  Ideally, such models should be as constrained
as possible from QCD.  Inequality (\ref{ineq1}) provides one
possible basis for such a constraint. While the left-hand side of the inequality will be given by the model, the right-hand side is tractable in lattice QCD.  The key point which allows such lattice simulations was recognized several years ago \cite{AlfKapWil99}
and is
simply that the integrand in the functional integral for $Z_I$ in
eq.~({\ref{ZI})  is positive definite and hence is amenable
to lattice studies.  Lattice studies have been done
both in the quenched approximation \cite{Kog1} and including
dynamical quarks \cite{Kog2}.  It is probably sensible to consider
the dynamical lattice studies as being rather preliminary as they have
been done on small lattices.  However, as the lattice calculations
improve, they may provide a strong constraint on models of QCD at finite baryon chemical potential.

Next consider inequality (\ref{ineq4}).  It is very similar to the bounds derived by Nussinov for a wide class of QCD-inspired potential  models  \cite{Nus83} and to the bound derived by  Nussinov and Sathiapalan  in large $N_c$ QCD \cite{NusSat85}.  The present result suffers in comparison to these results in that instead of bounding the nucleon mass by $N_c m_\pi /2 $ the bound is with respect to $M_N \ge \frac{N_c \, m_X}{  2 \, I_X}$ where $X$ is the state in QCD with lowest mass per unit isospin. In practice this disadvantage is fairly small---it so happens that in QCD the state with the lowest mass per unit isospin {\it is} the pion.  The present bound has a clear advantage over the the results in refs. \cite{Nus83,NusSat85} in that it holds for QCD itself and is not restricted to the large $N_c$ limit.  Similarly, the present bound is much stronger than   Weingarten's original bound \cite{Wei83} in that it holds for QCD with two degenerate light flavors and does not require the {\it ad hoc}
assumptions about the quark
propagator in the presence of a background field. In any event the inequality (\ref{ineq4}) is true phenomenologically.  There does exist a state in QCD for which $M_N > (3 \, M_X)/(2 \, I_X)$ namely the case where $X$ is the pion.
 Overall, one should view {\it any} bound on the nucleon mass in terms of other physical observables directly from a first principles treatment of QCD
as being highly non trivial.

The author thanks S. Nussinov for introducing him to the subject of QCD inequalities, for numerous illuminating discussions on the topic over the years and as a guide to the relevant literature. D.~Dakin is gratefully acknowledged for suggesting a number of improvements to the manuscript. This work was supported by the United States Department of Energy through grant DE-FGO2-93ER-40762.

\end{document}